# Terminal Adaptive Guidance for Autonomous Hypersonic Strike Weapons via Reinforcement Learning

Brian Gaudet*
*University of Arizona, 1127 E. Roger Way, Tucson Arizona, 85721*

Roberto Furfaro†
*University of Arizona, 1127 E. Roger Way, Tucson Arizona, 85721*

An adaptive guidance system suitable for the terminal phase trajectory of a hypersonic strike weapon is optimized using reinforcement meta learning. The guidance system maps observations directly to commanded bank angle, angle of attack, and sideslip angle rates. Importantly, the observations are directly measurable from radar seeker outputs with minimal processing. The optimization framework implements a shaping reward that minimizes the line of sight rotation rate, with a terminal reward given if the agent satisfies path constraints and meets terminal accuracy and speed criteria. We show that the guidance system can adapt to off-nominal flight conditions including perturbation of aerodynamic coefficient parameters, actuator failure scenarios, sensor scale factor errors, and actuator lag, while satisfying heating rate, dynamic pressure, and load path constraints, as well as a minimum impact speed constraint. We demonstrate precision strike capability against a maneuvering ground target and the ability to divert to a new target, the latter being important to maximize strike effectiveness for a group of hypersonic strike weapons. Moreover, we demonstrate a threat evasion strategy against interceptors with limited midcourse correction capability, where the hypersonic strike weapon implements multiple diverts to alternate targets, with the last divert to the actual target. Finally, we include preliminary results for an integrated guidance and control system in a six degrees-of-freedom environment.

## I. Introduction

There are multiple challenges associated with designing a guidance system for the terminal phase of a hypersonic strike weapon (HSW), where the goal is to impact a mobile and potentially maneuvering ground target. There are multiple objectives, including precision strike capability, maximizing impact speed to enhance lethality through kinetic energy, minimizing flight time to give the target less time to deploy defensive countermeasures, evading surface to air missiles, and satisfying path constraints on heating rate, dynamic pressure, and load. Moreover, the guidance system will likely be operating in a GPS denied environment where the target's location is bounded, but uncertain. Consequently, the HSW will need to autonomously meet mission objectives by selecting a target, potentially while coordinating with other strike weapons and evading target defensive systems. And since the target will have likely moved from its expected position, the guidance system needs to tolerate a reasonable range of initial heading errors. Once a target is selected, it can be tracked by a seeker. The guidance, navigation, and control system must then map seeker angles, their rate of change, range, closing speed, and rate gyro measurements to actuator commands. Finally, transitioning from 25 km altitude to sea level at hypersonic speeds creates a large heating rate, significant structural load, and an extremely large increase in dynamic pressure. This extreme aero-thermal environment can result in airframe deformation and control surface ablation, and the vehicle will experience off-nominal flight conditions that differ substantially from that assumed during optimization of the guidance system.

Previous work in terminal phase guidance for a hypersonic vehicle in a three degrees-of-freedom (3-DOF) environment include [1], which uses optimal control theory to generate a trajectory and compares inverted to non-inverted flight performance, with path constraints for heating rate, dynamic pressure, and load. In [2] the authors develop a guidance system that allows specification of impact direction, and in [3] the guidance law allows specification of impact time and angle. However, none of these papers measured performance under conditions of aerodynamic coefficient perturbation and considered a mobile target. Moreover, although a portion of the trajectory was at hypersonic speeds, impact speeds were in the supersonic range. In addition, these papers established performance using only a small number of specific cases rather than extensively testing the guidance system by running a large number of simulations with randomized initial conditions.

---

*Research Engineer, Department of Systems and Industrial Engineering, E-mail:briangaudet@arizona.edu
†Professor, Department of Systems and Industrial Engineering, Department of Aerospace and Mechanical Engineering. E-mail:robertof@arizona.edu

There has also been recent work in developing guidance and control systems for hypersonic terminal phase guidance in a 6-DOF environment. In [4] the authors develop a guidance and control system using separate inner and outer control loops and demonstrate impressive accuracy with aerodynamic coefficient perturbation, but do not consider mobile targets. Moreover, the loading constraint is set to 20 g, which given the high vehicle surface temperature at hypersonic speeds might induce structural failure. In [5] the authors develop an integrated guidance and control system that is provably stable using the Lyapunov method that can satisfy impact angle constraints, and demonstrate high accuracy with a slowly moving target (15 m/s) while considering aerodynamic coefficient perturbation. However, only a single heading error is taken into account, and the load constraint of 20 g might be too high to guarantee structural integrity. Finally, in [3] the authors develop a low order integrated guidance and control system and demonstrate accurate guidance in the presence of large perturbations to aerodynamic coefficients, but only consider a fixed 10 degree initial heading error, no target motion, and no path constraints. None of the published work we reviewed considered actuator lag, sensor scale factor errors [6], or actuator failure scenarios, and all assumed access to measurements (angle of attack, sideslip angle, flight path angle, heading angle) that would not be obtainable in practice. More importantly, previous work has not demonstrated divert maneuver or threat evasion capability during the terminal phase. This is an extremely useful capability for a guidance, navigation, and control system that can identify targets and evaluate their relative value. For example, an HSW may initially lock onto a target but may then identify a higher value target. Alternatively, the current target may be destroyed by another HSW in the strike group, requiring the HSW to acquire a new target.

In this work we do not attempt to solve the full problem as previously described. Instead, we model a single HSW and use a simplified 3-DOF environment, although preliminary 6-DOF results are given in Section IV.B. We assume that the HSW has a radar seeker that can identify and track a target at a range of 200 km, and the terminal phase begins at this distance at an altitude of 25 km and speed of 3000 m/s, with the full set of initial conditions given later in Section II.D, Table 1. The goal is to impact a maneuvering target at a maximum distance of 5 m from the target centroid at a terminal speed of at least 1700 m/s (Mach 5 at sea level). For large targets such as ships this would result in a direct impact of the HSW on the target. Given the vehicle's mass of 1361 kg, this would result in kinetic energy at impact of at least 2 GJ, or the equivalent of 1/2 ton of TNT in addition to any explosive payload. Importantly, we will also demonstrate divert capability, where the HSW must divert to a new target with a randomly generated position, velocity, and acceleration. Further, we demonstrate a threat evasion strategy that executes multiple divert maneuvers in a single trajectory. Although in the 3-DOF setting we cannot accurately model hypersonic flight, we attempt to capture the essence of the problem by simulating the vehicle in an environment that at the start of each episode randomly perturbs aerodynamic coefficients, atmospheric density, vehicle mass, and vehicle reference area. Moreover, we consider actuator failure scenarios, and take into account sensor scale factor errors and actuator lag. Finally, we impose path constraints on heating, dynamic pressure, and load. Importantly, the guidance law uses observations that would be directly measurable using a radar seeker. In the following, we develop a guidance system suitable for this scenario, with the guidance system implemented as a policy optimized using meta reinforcement learning [7–9] (meta-RL).

In the meta-RL framework, an agent instantiating the policy learns through episodic simulated experience over an ensemble of environments covering the expected distribution of mission scenarios, aerodynamic regimes, sensor and actuator degradation, variation in system time constants, and other factors. The policy is implemented as a deep neural network parameterized by $\theta$ that maps observations to actions $\mathbf{u} = \pi_\theta(\mathbf{o})$, and is optimized using a customized version of proximal policy optimization (PPO) [10]. Adaptation is achieved by including a recurrent network layer [11] with hidden state $\mathbf{h}$ in both the policy and value function networks. Maximizing the PPO objective function requires learning hidden layer parameters $\theta_h$ that result in $\mathbf{h}$ evolving in response to the history of $\mathbf{o}$ and $\mathbf{u}$ in a manner that facilitates fast adaptation to an environment sampled from the ensemble, and generalization to novel environments. The deployed policy will then adapt to off-nominal conditions during flight. Importantly, the network parameters remain fixed during deployment with adaptation occurring through the evolution of $\mathbf{h}$. Although reinforcement learning has been previously applied to aerospace guidance, navigation, and control applications [12], [13], [14], [15], [16], and [17], to our knowledge, this is the first published work that uses meta-RL to optimize a guidance system suitable for the terminal phase of a HSW, and the first published work to demonstrate divert capability and evasive maneuvers in this application.

## II. Problem Formulation

### A. Equations of Motion

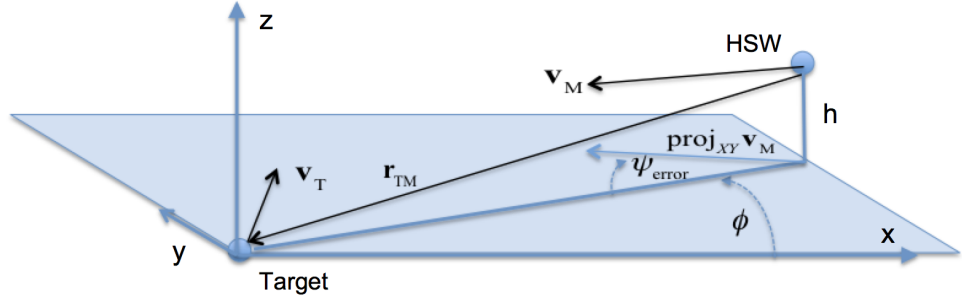

**Fig. 1  Engagement Geometry**

Figure 1 illustrates the engagement geometry, with both the target and HSW in their initial state. Let $\mathbf{r}_M$ and $\mathbf{r}_T$ denote the position of the HSW and target, and $\mathbf{v}_M$ and $\mathbf{v}_T$ be the velocity of the HSW and target. The initial position of the HSW $\mathbf{r}_{M_{init}}$ can be described in terms of an initial range to target $r_{init}$, altitude $h_{init}$, and azimuth angle $\phi_{init}$ as shown in Eqs. (1a) and (1b). The initial target position $\mathbf{r}_T$ is at the origin of the coordinate system, and $\mathbf{r}_{TM} = \mathbf{r}_T - \mathbf{r}_M$ is the vector from the HSW to the target, with $r = \|\mathbf{r}_{TM}\|$ the distance between the HSW and target.

$$\theta_{init} = \arcsin \frac{h_{init}}{r_{init}} \tag{1a}$$

$$\mathbf{r}_{M_{init}} = \begin{bmatrix} r_{init} \cos\theta_{init} \cos\phi_{init} & r_{init} \cos\theta_{init} \sin\phi_{init} & h_{init} \end{bmatrix} \tag{1b}$$

It is convenient to use spherical coordinates to describe the evolution of the HSW velocity vector in the system dynamics. Let $\mathbf{v}_M = \begin{bmatrix} v_{M_x} & v_{M_y} & v_{M_z} \end{bmatrix}$ be the HSW velocity vector in the engagement coordinate system given in Fig. 1. Then the HSW speed $V$, flight path angle $\gamma$, and heading angle $\psi$ are as shown in Eqs. (2a), with the inverse mapping shown in Eq. (2b). In the following we will denote the mapping from Cartesian to spherical velocity as $c2s(\cdot)$ and the inverse mapping as $s2c(\cdot)$,

$$\begin{bmatrix} V & \gamma & \psi \end{bmatrix} = \begin{bmatrix} \|\mathbf{v}_M\| & \arcsin \frac{v_{M_z}}{V} & \arctan2(v_{M_y}, v_{M_x}) \end{bmatrix} \tag{2a}$$

$$\mathbf{v}_M = \begin{bmatrix} V\cos\gamma\cos\psi & V\cos\gamma\sin\psi & V\sin\gamma \end{bmatrix} \tag{2b}$$

We assume that the HSW has no knowledge of the target's velocity vector, and therefore do not define an ideal HSW heading as being on a collision triangle with the target, but instead define an ideal heading $\psi_{ideal}$ as the projection of $\mathbf{r}_{TM_{init}}$ onto the x-y plane, and the initial heading as $\psi_{init} = \psi_{ideal} + \psi_{error}$, where the heading error $\psi_{error}$ accounts for uncertainty in the approach phase trajectory and potential target movement during the approach phase. The initial HSW velocity vector $\mathbf{v}_{M_{init}}$ is then as shown in Eq. (3a), where $\gamma_{init}$ is the initial flight path angle.

$$\mathbf{v}_{M_{init}} = s2c\left(\begin{bmatrix} V_{init} & \gamma_{init} & \psi_{init} \end{bmatrix}\right) \tag{3a}$$

The HSW state then evolves according to Eqs. (4a) through (4h), where after each integration step the HSW velocity vector is obtained as $\mathbf{v}_M = s2c\left(\begin{bmatrix} V & \gamma & \psi \end{bmatrix}\right)$. Here $L$, $D$, and $Y$ are the lift, drag and side forces, $\alpha$, $\beta$, and $\nu$ are the angle of attack, sideslip angle, and bank angle, the HSW mass is $m$, and $g = 9.81$ m/s$^2$ is the acceleration due to gravity. $\Delta\alpha$, $\Delta\beta$, and $\Delta\nu$ are derived from the commanded changes to angle of attack, sideslip angle, and bank angle, as described in Section II.B. The equations of motion are integrated using fourth order Runge-Kutta integration with a 0.1 s timestep, with the timestep reduced by a factor of 300 when the HSW is within 1200 m of the target. Thus, the best possible simulated average accuracy at a terminal speed of 2500 m/s is 1.7 m.

$$\begin{bmatrix} V & \gamma & \psi \end{bmatrix} = c2s(\mathbf{v}_M) \tag{4a}$$

$$\dot{V} = -\frac{D}{m} - g \sin \psi \tag{4b}$$

$$\dot{\gamma} = \frac{L \cos \nu - Y \sin \nu}{mV} - \frac{g \cos \gamma}{V} \tag{4c}$$

$$\dot{\psi} = \frac{L \sin \nu + Y \cos \nu}{mV \cos \gamma} \tag{4d}$$

$$\dot{\mathbf{r}}_M = s2c\left(\begin{bmatrix} V & \gamma & \psi \end{bmatrix}\right) \tag{4e}$$

$$\dot{\alpha} = \Delta\alpha \tag{4f}$$

$$\dot{\nu} = \Delta\nu \tag{4g}$$

$$\dot{\beta} = \Delta\beta \tag{4h}$$

The drag, lift, and side forces are calculated as shown in Eqs. (5a) through (5c), where $S_{\text{ref}}$ is the aerodynamic reference area, and the atmospheric density $\rho$ is calculated using the exponential atmosphere model $\rho = \rho_0 e^{-(h)/h_s}$, where $\rho_0 = 1.225$ kg/m$^3$ is the density at sea level, and $h_o = 7018.00344$ m is the density scale-height.

$$D = \frac{\rho V^2 S_{\text{ref}} C_D}{2} \tag{5a}$$

$$L = \frac{\rho V^2 S_{\text{ref}} C_L}{2} \tag{5b}$$

$$Y = \frac{\rho V^2 S_{\text{ref}} C_Y}{2} \tag{5c}$$

Target motion is modeled as shown in Eqs. (6a) and (6b), where $\mathbf{a}_T$ is the target acceleration. Let $\mathcal{U}(a, b, d)$ denote a $d$ dimensional uniformly distributed random variable bounded by $(a, b)$. At the start of each episode, the initial target velocity is randomly generated as $\mathbf{v}_{T_{\text{init}}} = V_{T_{\text{init}}} \frac{\zeta_v}{\|\zeta_v\|}$, where $\zeta_v = \mathcal{U}(-1, 1, 3)$ and $V_{T_{\text{init}}} = \mathcal{U}(0, 30, 1)$ m/s. Similarly, the initial target acceleration is given by $\mathbf{a}_{T_{\text{init}}} = A_{T_{\text{init}}} \frac{\zeta_a}{\|\zeta_a\|}$, where $\zeta_a = \mathcal{U}(-1, 1, 3)$ and $A_{T_{\text{init}}} = \mathcal{U}(0, 0.5, 1)$ m/s$^2$. The target speed is clipped to a maximum value of 30 m/s$^2$.

$$\dot{\mathbf{r}}_T = \mathbf{v}_T \tag{6a}$$

$$\dot{\mathbf{v}}_T = \mathbf{a}_T \tag{6b}$$

Actuator delay is then modeled by integrating Eqs. (7a) through (7c), where $\tau_{\text{ctrl}}$, is the actuator time constant, and $\Delta\nu$, $\Delta\alpha$, and $\Delta\beta$ are the control inputs used in Eqs. (4f) through (4h).

$$\dot{\Delta\nu} = \frac{\Delta\nu_{\text{AF}} - \Delta\nu}{\tau_{\text{ctrl}}} \tag{7a}$$

$$\dot{\Delta\alpha} = \frac{\Delta\alpha_{\text{AF}} - \Delta\alpha}{\tau_{\text{ctrl}}} \tag{7b}$$

$$\dot{\Delta\beta} = \frac{\Delta\beta_{\text{AF}} - \Delta\beta}{\tau_{\text{ctrl}}} \tag{7c}$$

## B. Observation Model and Control

Our guidance policy is based on proportional navigation [18], which attempts to minimize the rotation rate of the line of sight vector to the target. The guidance policy $\pi$ will have access to an observation $\mathbf{o}$ that is a function of the ground truth vehicle state as computed in Section II.A. The un-biased observation is given as shown in Eq. (8a), where $\lambda = \dfrac{\mathbf{r}_{TM}}{\|\mathbf{r}_{TM}\|}$ is the line of sight unit vector, $\mathbf{\Omega} = \dfrac{\mathbf{r}_{TM} \times \mathbf{v}_{TM}}{\mathbf{r}_{TM} \cdot \mathbf{r}_{TM}}$ is the line of sight rotation vector [18], $r$ is the distance between the HSW and target, and $v_c = -\dfrac{\mathbf{r}_{TM} \cdot \mathbf{v}_{TM}}{\|\mathbf{r}_{TM}\|}$ is the closing velocity. $\tilde{\mathbf{o}}$ is then biased by a sensor scale factor error $\epsilon$ as shown in Eq. (8b), where $\mathbf{o} \in \mathcal{R}^n$. The sensor scale factor error is used to model radome refraction and other sensor biases. Note that in this work we do not accurately model the parasitic attitude loop [19]. Importantly, $\lambda$, $\mathbf{\Omega}$, $v_c$, and $r$ can be obtained by a radar seeker. Note that the fact that $\alpha$, $\beta$, and $\nu$ are not measurable at hypersonic speeds is not an issue, these observables are only included because the control consists of $\Delta \nu$, $\Delta \alpha$, and $\Delta \beta$. In an actual implementation, control would be through control surface deflection rates, and only the resulting deflections would need to be included in the observation, which we demonstrate in Section IV.B. Note that $\lambda$ and $\mathbf{\Omega}$ are not directly observable. Consequently, in a 6-DOF model, $\lambda$ would be replaced by the seeker azimuth and elevation angles, and $\mathbf{\Omega}$ would be replaced by first differences of the smoothed seeker angles, see [12] or [17] for possible implementations with a strapdown seeker.

$$\tilde{\mathbf{o}} = \begin{bmatrix} \lambda & \mathbf{\Omega} & v_c & r & \alpha & \beta & \nu \end{bmatrix} \tag{8a}$$

$$\mathbf{o} = \tilde{\mathbf{o}}(1 + \mathcal{U}(-\epsilon_{\text{obs}}, \epsilon_{\text{obs}}, n)) \tag{8b}$$

The output of the guidance policy $\mathbf{u} = \pi(\mathbf{o}) \in \mathbb{R}^3$ can be split into three components, commanded changes to the bank angle $\mathbf{u}[0]$, angle of attack $\mathbf{u}[1]$, and sideslip angle $\mathbf{u}[2]$. These control inputs are then multiplied by the maximum allowed rate of change for bank angle $\Delta \nu_{\max}$, angle of attack $\Delta \alpha_{\max}$, and sideslip angle $\Delta \beta_{\max}$, and then clipped to fall within the allowable limits, i.e., the commanded rate of change to bank angle is clipped to fall between $\pm \Delta \nu_{\max}$. We model actuator failure as follows. At the start of each episode, a failure occurs in each control channel with probability $p_{\text{fail}}$. Failure is modeled by multiplying the commanded control (i.e., $\Delta \alpha_{\max}$) by a scale factor $\epsilon_{\text{ctrl}} \in \mathcal{R}^2$ that is uniformly generated at the start of the episode. Finally, we add Gaussian actuator noise with standard deviation $\sigma_{\text{ctrl}}$ to each control channel.

## C. Divert and Evasive Maneuver Scenarios

With a phased array radar and target discrimination capability, the HSW GN&C system can track a primary target while simultaneously tracking several alternate targets. If the primary target is destroyed by another HSW in the strike group, or a higher value target is identified, the HSW should divert to an alternate target that allows a feasible strike. We will model this by introducing a random divert maneuver into each episode. Specifically, let $r_{\text{divert}}$ be the range to target at which the divert is executed. Then at the start of each episode, with probability $p_{\text{divert}}$ a divert maneuver is constructed by sampling $r_{\text{divert}}$ from a uniform distribution $r_{\text{divert}} = \mathcal{U}(30 \text{ km}, 150 \text{ km}, 1)$. When the actual range to target $r$ falls below $r_{\text{divert}}$, a divert distance is calculated as $\Delta_{\text{divert}} = \begin{bmatrix} \mathcal{U}(-0.05 \, r_{\text{divert}}, 0.05 \, r_{\text{divert}}, 2) & 0 \end{bmatrix}$, $\mathbf{r}_T$ is set to $\mathbf{r}_T + \Delta_{\text{divert}}$, and $\mathbf{v}_T$ and $\mathbf{a}_T$ are randomly reset as described at the end of Section II.A. Thus, the target location, speed, and acceleration are instantaneously reset to random values, as would be the case if a new target were acquired during the engagement.

The HSW can also implement multiple diverts to alternate, or even non-existent, targets, with the final divert putting the HSW on a course to the actual target. If the interceptors launched by the target have limited midcourse adjustment capability, then it is possible that at the point the interceptor begins its homing phase, the evasive maneuvers executed by the HSW will have created a large enough heading error to make interception infeasible. We model this evasive behavior by having the HSW execute multiple random diverts, switching to alternate targets $\Delta_{\text{divert}}$ from the last target, with a minimum distance of 30km between diverts. These diverts start at a distance of 150km to the target and end at a distance of 25km to the target.

## D. Vehicle Model and Mission

We use the generic hypersonic vehicle (GHV) described in [20, 21], but with the aerodynamic coefficient model used in [4], where $C_L$ and $C_D$ are functions of $\alpha$ and Mach number $M$, and $C_Y$ a function of $\alpha$, $M$, and $\beta$, as shown in Eqs. (9a) through (9c). Although in [20] the GHV is assumed to be powered, in this work we assume that the GHV is unpowered and deployed in a boost-glide scenario. The mass $m$ and reference area $S_{\text{ref}}$ are scaled to 1361kg and 3.347m$^2$ (without changing the ratio of mass to reference area) to give a more realistic size for a hypersonic strike weapon.

$$C_L = \begin{bmatrix} -0.081929 \\ 0.0470142 \\ -0.00919 \\ 0.000774 \\ -0.0000293 \\ 0.000000412 \end{bmatrix}^T \begin{bmatrix} M^0 \\ M^1 \\ M^2 \\ M^3 \\ M^4 \\ M^5 \end{bmatrix} + \begin{bmatrix} 1.07727 - 0.0265M \\ -0.49898 + 0.0019M^2 \\ 0.76741107 \\ -4.21373565 \\ 8.02706009 \end{bmatrix}^T \begin{bmatrix} \alpha^1 \\ \alpha^2 \\ \alpha^3 \\ \alpha^4 \\ \alpha^5 \end{bmatrix} \quad (9a)$$

$$C_D = \begin{bmatrix} 0.08883096 \\ -0.03339562 \\ 0.005044728 \\ -0.0003658 \\ 0.00001274 \\ -0.00000017 \end{bmatrix}^T \begin{bmatrix} M^0 \\ M^1 \\ M^2 \\ M^3 \\ M^4 \\ M^5 \end{bmatrix} + \begin{bmatrix} 0.183 - 0.00716M \\ -3.587 + 0.0005M^2 \\ 59.71887625 \\ -321.68800332 \\ 603.01745298 \end{bmatrix}^T \begin{bmatrix} \alpha^1 \\ \alpha^2 \\ \alpha^3 \\ \alpha^4 \\ \alpha^5 \end{bmatrix} \quad (9b)$$

$$C_Y = \begin{bmatrix} -0.29253 \\ 0.054822 \\ -0.0043203 \\ 0.00015495 \\ -0.0000020829 \end{bmatrix}^T \begin{bmatrix} M^1 \\ M^2 \\ M^3 \\ M^4 \\ M^5 \end{bmatrix} \beta + \begin{bmatrix} 0.16502903 - 0.01658312M \\ 2.41401 + 0.01516821M^2 \\ -70.3554194 \\ 303.723 - 0.2228107M^2 \\ -321.59490071 \end{bmatrix}^T \begin{bmatrix} \alpha^1 \\ \alpha^2 \\ \alpha^3 \\ \alpha^4 \\ \alpha^5 \end{bmatrix} \beta \quad (9c)$$

The vehicle is assumed to have a carbon aeroshell with a melting point of 3500 K, and that maximum temperature occurs at the vehicle nose. We use the equations relating heating rate to wall temperature at the nose given in [22] for the HTV-2 vehicle, which uses a leading edge radius of $r_n = 0.034$m. Here we assume that the GHV, scaled as previously described, has a similar leading edge radius. The equations governing heating rate $\dot{Q}$ and wall temperature $T_w$ are taken from [22] and listed in Eqs. (10a) and (10b), where the wall enthalpy is given by $h_w = 1000T_w$ J/kg, the stagnation enthalpy by $h_o = \frac{V^2}{2} + 2.3 \times 10^5$ J/kg, $\epsilon = 0.85$ is the surface emissivity, and $\sigma_{\text{SB}}$ the Stefan-Boltzmann constant. Note that the equations in [22] were formulated based on the work in [23].

$$\dot{Q} = \frac{1.83 \times 10^{-4}}{\sqrt{r_n}} \left(1 - \frac{h_w}{h_o}\right) \sqrt{\rho} V^3 \quad (10a)$$

$$T_w = \left(\frac{\dot{Q}}{\epsilon \sigma_{\text{SB}}}\right)^{1/4} \quad (10b)$$

Reference [23] suggests iterating between Eqs. (10a) and (10b) to converge on heating rate and wall temperature. However, early in learning, the agent can induce trajectories that result in $h_w > h_o$. Thus, rather than develop a heating model that would work over a much wider range of conditions than would be encountered in a sensible trajectory, we approximate the heating rate by assuming that $h_w = 0.50h_o$, as shown in Eq. (11a). At an altitude of 3 km and a speed of 2500 m/s, this corresponds to a heating rate of $8.5 \times 10^6$ kW/m$^2$, and a wall temperature of 3650 K, making it likely that some form of active cooling or active flow field modification would be required. Note that we make the same assumptions as [22] and assume that the vehicle aeroshell is at thermal equilibrium, and therefore the vehicle aeroshell temperature is equal to the wall temperature. We also impose a dynamic pressure and load constraint, with the three path constraints listed in Eqs. (11a) through (11e), where $\dot{Q}_{\max} = 9000$kW/m$^2$, $q_{\max} = 4000$kg/m-s$^2$, $n_{\max} = 15$ m/s$^2$, and $[\mathbf{C}_{\text{BW}}(\alpha, \beta)]$ is the direction cosine matrix mapping from the wind to body frame.

$$\dot{Q} = \frac{1.83 \times 10^{-4}}{\sqrt{r_n}} (0.50) \sqrt{\rho} V^3 \tag{11a}$$

$$q = \frac{1}{2}\rho V^2 \leq q_{max} \tag{11b}$$

$$\mathbf{F}^{wind} = \begin{bmatrix} D & Y & L \end{bmatrix} \tag{11c}$$

$$\mathbf{F}^{body} = [\mathbf{C}_{BW}(\alpha, \beta)] \mathbf{F}^{wind} \tag{11d}$$

$$n = \left\| \begin{bmatrix} \mathbf{F}_y^{body} & \mathbf{F}_z^{body} \end{bmatrix} \right\| \leq n_{max} \tag{11e}$$

In this work we assume that the HSW has a radar seeker that can identify and track a target at a range of 200 km, and the terminal phase begins at this distance at an altitude of 25 km and speed of 3000 m/s. The full set of initial conditions are given in Table 1. In recent work with approach phase guidance [15], we have achieved terminal phase accuracy consistent with these initial conditions. The goal is to impact the target at a maximum distance of 5 m from the target centroid at a terminal speed of at least 1700 m/s (Mach 5 at sea level); for large targets such as ships this would result in the HSW impacting the target.

**Table 1   Initial Conditions**

| Parameter | min | max |
| --- | --- | --- |
| Range $d_{init}$ (km) | 200 | 200 |
| Azimuth $\phi$ (degrees) | -10 | 10 |
| Heading Error $\psi_{error}$ (degrees) | -10 | 10 |
| Altitude $h_{init}$ (km) | 24.8 | 25.2 |
| Speed $V_{init}$ (m/s) | 2900 | 3100 |
| Flight Path Angle $\gamma_{init}$ (degrees) | -5 | 0 |
| Angle of Attack $\alpha_{init}$ (degrees) | 1 | 3 |
| Bank Angle $\sigma_{init}$ (degrees) | -2 | 2 |
| Sideslip Angle $\sigma_{init}$ (degrees) | -2 | 2 |

The vehicle parameters are given in Table 2, and Table 3 shows the range over which selected parameters in the dynamics model are perturbed during optimization. Each parameter $\mathbf{p} \in \mathbb{R}^d$ has its value randomly set at the start of an episode as $\mathbf{p} = \mathcal{U}(-\nu, \nu, 1)$, where $\nu$ is the bound in the "Value" column of Table 3. This perturbation is meant to capture differences between the dynamics model used for optimization and the actual dynamics encountered in deployment. Finally, Table 4 tabulates the target model parameters. The guidance system maps observations to actions at a guidance frequency of 5 Hz, i.e., the update occurs every 0.2 seconds.

Table 2  Vehicle Parameters

| Parameter | Value |
|---|---|
| Mass $m$ (kg) | 1361 |
| Area Reference $S_{ref}$ (m$^2$) | 3.347 |
| Guidance Period (s) | 0.5 |
| Bank Angle Limits $\sigma_{max}$ (degrees) | (-180,180) |
| Angle of Attack Limits $\alpha_{max}$ (degrees) | (0,12) |
| Sideslip Angle Limits $\beta_{max}$ (degrees) | (-12,12) |
| Angle of Attack Rate Limit $\Delta\alpha_{max}$ (degrees / s) | 4.0 |
| Sideslip Angle Rate Limit $\Delta\beta_{max}$ (degrees / s) | 4.0 |
| Bank Angle Rate Limit $\Delta\sigma_{max}$ (degrees / s) | 10.0 |
| Actuator lag time constant $\tau_{ctrl}$ (s) | 0.1 |
| Actuator Noise Standard Deviation $\sigma_{ctrl}$ (degrees / s) | 0.1 |
| Actuator Failure Bias Range $\epsilon_{ctrl}$ | (-0.3, 0.0) |
| Actuator Failure Probability $p_{fail}$ | 0.5 |
| Sensor Scale Factor Error $\epsilon_{obs}$ | 0.005 |

Table 3  Dynamics Model Variation

| Parameter | Value |
|---|---|
| Aerodynamic lift coefficient $C_L$ variation | 10% |
| Aerodynamic sideforce coefficient $C_Y$ variation | 10% |
| Aerodynamic drag coefficient $C_D$ variation | 10% |
| Atmospheric density $\rho$ variation | 10% |

Table 4  Target Model

| Parameter | Value |
|---|---|
| Maximum Speed (m/s) | 30 |
| Maximum Acceleration (m/s$^2$) | 0.5 |
| Divert Probability $p_{divert}$ | 0.5 |

## III. Methods

### A. Background: Reinforcement Learning Framework

In the reinforcement learning framework, an agent learns through episodic interaction with an environment how to successfully complete a task using a policy that maps observations to actions. The environment initializes an episode by randomly generating a ground truth state, mapping this state to an observation, and passing the observation to the agent. The agent uses this observation to generate an action that is sent to the environment; the environment then uses the action and the current ground truth state to generate the next state and a scalar reward signal. The reward and the observation corresponding to the next state are then passed to the agent. The process repeats until the environment terminates the episode, with the termination signaled to the agent via a done signal. Trajectories collected over a set of episodes (referred to as rollouts) are collected during interaction between the agent and environment, and used to update the policy and value functions. The interface between agent and environment is depicted in Fig. 2.

A Markov Decision Process (MDP) is an abstraction of the environment, which in a continuous state and action space, can be represented by a state space $\mathcal{S}$, an action space $\mathcal{A}$, a state transition distribution $\mathcal{P}(\mathbf{x}_{t+1}|\mathbf{x}_t, \mathbf{u}_t)$, and a reward function $r = \mathcal{R}(\mathbf{x}_t, \mathbf{u}_t)$, where $\mathbf{x} \in \mathcal{S}$ and $\mathbf{u} \in \mathcal{A}$, and $r$ is a scalar reward signal. We can also define a partially observable MDP (POMDP), where the state $\mathbf{x}$ becomes a hidden state, generating an observation $\mathbf{o}$ using an observation function $O(\mathbf{x})$ that maps states to observations. The POMDP formulation is useful when the observation consists of sensor outputs. In the following, we will refer to both fully observable and partially observable environments as POMDPs, as an MDP can be considered a POMDP with an identity function mapping states to observations.

Meta-RL differs from generic reinforcement learning in that the agent learns over an ensemble of POMPDs. These POMDPs can include different environmental dynamics, aerodynamic coefficients, actuator failure scenarios, mass and inertia tensor variation, and varying amounts of sensor distortion. Optimization within the meta-RL framework results in an agent that can quickly adapt to novel POMDPs, often with just a few steps of interaction with the environment. There are multiple approaches to implementing meta-RL. In [24], the authors design the objective function to explicitly make the model parameters transfer well to new tasks. In [7], the authors demonstrate state of the art performance using temporal convolutions with soft attention. And in [8], the authors use a hierarchy of policies to achieve meta-RL. In this work, we use an approach similar to [9] using a recurrent policy and value function. Note that it is possible to train over a wide range of POMDPs using a non-meta RL algorithm. Although such an approach typically results in a robust policy, the policy cannot adapt in real time to novel environments. In this work, we implement meta-RL using proximal policy optimization (PPO) [10] with both the policy and value function implementing recurrent layers in their networks. After training, although the recurrent policy's network weights are frozen, the hidden state will continue to evolve in response to a sequence of observations and actions, thus making the policy adaptive. In contrast, a policy without a recurrent layer has behavior that is fixed by the network parameters at test time.

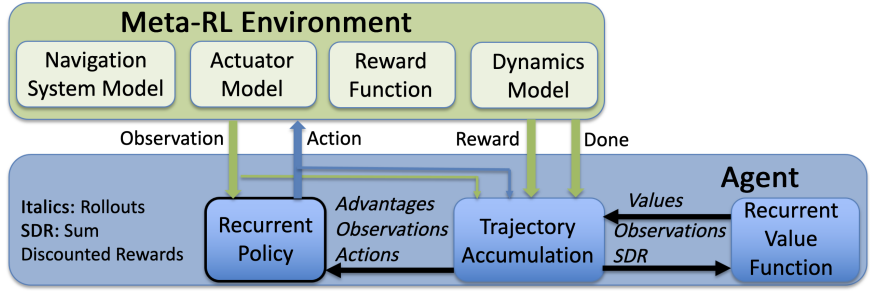

**Fig. 2** Environment-Agent Interface

The PPO algorithm used in this work is a policy gradient algorithm which has demonstrated state-of-the-art performance for many reinforcement learning benchmark problems. PPO approximates the Trust Region Policy Optimization method [25] by accounting for the policy adjustment constraint with a clipped objective function. The objective function used with PPO can be expressed in terms of the probability ratio $p_k(\boldsymbol{\theta})$ given by,

$$p_k(\boldsymbol{\theta}) = \frac{\pi_{\boldsymbol{\theta}}(\mathbf{u}_k|\mathbf{o}_k)}{\pi_{\boldsymbol{\theta}_{\text{old}}}(\mathbf{u}_k|\mathbf{o}_k)} \tag{12}$$

The PPO objective function is shown in Equations (13a) through (13c). The general idea is to create two surrogate objectives, the first being the probability ratio $p_k(\boldsymbol{\theta})$ multiplied by the advantages $A_{\mathbf{w}}^{\pi}(\mathbf{o}_k, \mathbf{u}_k)$ (see Eq. (14)), and the second a clipped (using clipping parameter $\epsilon$) version of $p_k(\boldsymbol{\theta})$ multiplied by $A_{\mathbf{w}}^{\pi}(\mathbf{o}_k, \mathbf{u}_k)$. The objective to be maximized $J(\boldsymbol{\theta})$ is then the expectation under the trajectories induced by the policy of the lesser of these two surrogate objectives.

$$\text{obj1} = p_k(\boldsymbol{\theta}) A_{\mathbf{w}}^{\pi}(\mathbf{o}_k, \mathbf{u}_k) \tag{13a}$$

$$\text{obj2} = \text{clip}(p_k(\boldsymbol{\theta}) A_{\mathbf{w}}^{\pi}(\mathbf{o}_k, \mathbf{u}_k), 1-\epsilon, 1+\epsilon) \tag{13b}$$

$$J(\boldsymbol{\theta}) = \mathbb{E}_{p(\tau)}[\min(\text{obj1}, \text{obj2})] \tag{13c}$$

This clipped objective function has been shown to maintain a bounded Kullback-Leibler (KL) divergence [26] with respect to the policy distributions between updates, which aids convergence by ensuring that the policy does not change drastically between updates. Our implementation of PPO uses an approximation to the advantage function that is the difference between the empirical return and a state value function baseline, as shown in Equation 14, where $\gamma$ is a discount rate and $r$ the reward function, described in Section III.B.

$$A_{\mathbf{w}}^{\pi}(\mathbf{x}_k, \mathbf{u}_k) = \left[\sum_{\ell=k}^{T} \gamma^{\ell-k} r(\mathbf{o}_\ell, \mathbf{u}_\ell)\right] - V_{\mathbf{w}}^{\pi}(\mathbf{x}_k) \tag{14}$$

Here the value function $V_{\mathbf{w}}^{\pi}$ is learned using the cost function given by

$$L(\mathbf{w}) = \frac{1}{2M} \sum_{i=1}^{M} \left( V_{\mathbf{w}}^{\pi}(\mathbf{o}_k^i) - \left[ \sum_{\ell=k}^{T} \gamma^{\ell-k} r(\mathbf{u}_\ell^i, \mathbf{o}_\ell^i) \right] \right)^2 \quad (15)$$

In practice, policy gradient algorithms update the policy using a batch of trajectories (roll-outs) collected by interaction with the environment. Each trajectory is associated with a single episode, with a sample from a trajectory collected at step $k$ consisting of observation $\mathbf{o}_k$, action $\mathbf{u}_k$, and reward $r_k(\mathbf{o}_k, \mathbf{u}_k)$. Finally, gradient ascent is performed on $\theta$ and gradient descent on $\mathbf{w}$ and update equations are given by

$$\mathbf{w}^+ = \mathbf{w}^- - \beta_{\mathbf{w}} \nabla_{\mathbf{w}} L(\mathbf{w})|_{\mathbf{w}=\mathbf{w}^-} \quad (16)$$
$$\theta^+ = \theta^- + \beta_\theta \nabla_\theta J(\theta)|_{\theta=\theta^-} \quad (17)$$

where $\beta_{\mathbf{w}}$ and $\beta_\theta$ are the learning rates for the value function, $V_{\mathbf{w}}^{\pi}(\mathbf{o}_k)$, and policy, $\pi_\theta(\mathbf{u}_k|\mathbf{o}_k)$, respectively.

In our implementation of PPO, we adaptively scale the observations and servo both $\epsilon$ and the learning rate to target a KL divergence of 0.001.

### B. Meta-RL Problem Formulation

In this terminal phase hypersonic guidance application an episode terminates when the closing velocity $v_c$ turns negative, the time of flight exceeds 120 s, or the heating rate, dynamic pressure, or load constraints are violated. Rollouts are collected over 60 episodes of interaction between the agent and environment and used to update the policy and value functions. The agent observation $\mathbf{o}$ was given in Section II.B, and the processing of the agent action $\mathbf{u} \in \mathbb{R}^3$ was discussed in Section II.B. The reward function is shown below in Equations (18a) through (18d). $r_{\text{shaping}}$ is a shaping reward given at each step in an episode. These shaping rewards take the form of a Gaussian-like function of the norm of the line of sight rotation rate $\mathbf{\Omega}$. $r_{\text{ctrl}}$ is a control effort penalty, again given at each step in an episode, and $r_{\text{bonus}}$ is a bonus given at the end of an episode if certain conditions are met. Since rewards are discounted, this terminal reward encourages actions that minimize flight time. Importantly, the current episode is terminated if a constraint is violated, in which case the stream of positive shaping rewards is terminated, and the agent does not receive the terminal reward. It turns out that this is enough to incentivize the agent to meet the constraints, and it is not necessary to give the agent a negative reward when a constraint is violated. We use $\alpha = 1$, $\beta = -0.01$, $\epsilon = 20$, $r_{\text{lim}} = 50$ m, $v_{\text{lim}} = 1700$ m/s, $\sigma_\Omega = 0.05$. We use a discount rate of 0.90 for shaping rewards and 0.995 for the terminal reward.

$$r_{\text{shaping}} = \alpha \exp\left(\frac{-\|\mathbf{\Omega}\|^2}{\sigma_\Omega^2}\right) \quad (18a)$$

$$r_{\text{ctrl}} = \left\| \left[ \frac{\Delta\alpha_{\text{cmd}}}{\Delta\alpha_{\text{max}}} \quad \frac{\Delta\beta_{\text{cmd}}}{\Delta\beta_{\text{max}}} \quad \frac{\Delta\sigma_{\text{cmd}}}{\Delta\sigma_{\text{max}}} \right] \right\| \quad (18b)$$

$$r_{\text{bonus}} = \begin{cases} \epsilon, & \text{if } h < 0 \text{ and } r < r_{\text{lim}} \text{ and } \|\mathbf{v}_M > v_{\text{lim}}\| \text{ and done} \\ 0, & \text{otherwise} \end{cases} \quad (18c)$$

$$r = r_{\text{shaping}} + r_{\text{ctrl}} + r_{\text{bonus}} \quad (18d)$$

The policy and value functions are implemented using four layer neural networks with tanh activations on each hidden layer. Layer 2 for the policy and value function is a recurrent layer implemented using gated recurrent units [11]. The network architectures are as shown in Table 5, where $n_{\text{hi}}$ is the number of units in layer $i$, obs_dim is the observation dimension, and act_dim is the action dimension. The policy and value functions are periodically updated during optimization after accumulating trajectory rollouts of 60 simulated episodes.

### C. Optimization

Optimization uses the initial conditions and vehicle parameters given in Section II.D. The agent quickly learns to satisfy the path constraints, and it is possible that the path constraints could have been tightened. Once the agent learns to satisfy the constraints, the agent adjusts its policy to maximize both shaping and terminal rewards while continuing to satisfy constraints. Learning curves are given in Figures 3 through 5. Fig. 3 plots the reward history (sum of shaping and terminal rewards), with the mean ("Mean R"), mean minus 1 standard deviation ("SD R"), and minimum ("Min R") rewards plotted on the primary y-axis and the mean and maximum number of steps per episode plotted on the secondary y-axis. Similarly, Fig. 4 plots terminal miss distance statistics. Finally, Fig. 5 plots the terminal reward history. These statistics are computed over a batch of rollouts (60 episodes). It is worth noting that early in optimization the agent explores trajectories that result in inverted flight, but eventually converges on a solution with more limited bank angles. This could be due to the impact of control effort penalties, or perhaps the non-inverted trajectories gave better performance.

Table 5  Policy and Value Function network architecture

|  | Policy Network | | Value Network | |
| --- | ---: | :---: | ---: | :---: |
| Layer | # units | activation | # units | activation |
| hidden 1 | $10 * \text{obs\_dim}$ | tanh | $10 * \text{obs\_dim}$ | tanh |
| hidden 2 | $\sqrt{n_{h1} * n_{h3}}$ | tanh | $\sqrt{n_{h1} * n_{h3}}$ | tanh |
| hidden 3 | $10 * \text{act\_dim}$ | tanh | 5 | tanh |
| output | act_dim | linear | 1 | linear |

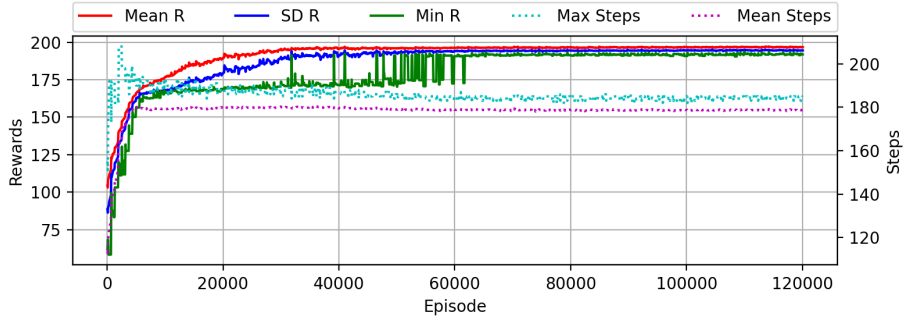

Fig. 3   Optimization Reward History

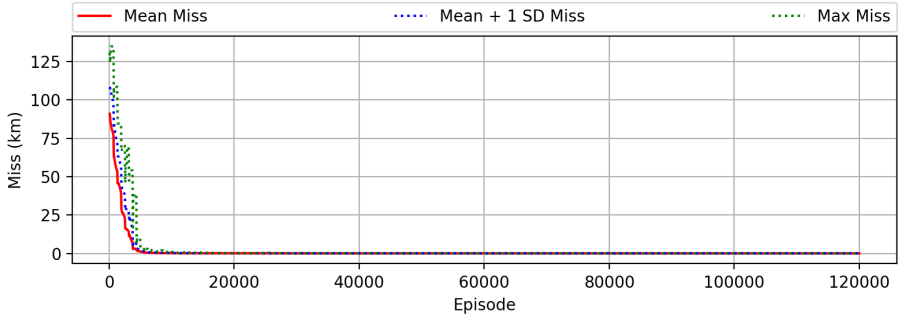

Fig. 4   Optimization Miss Distance History

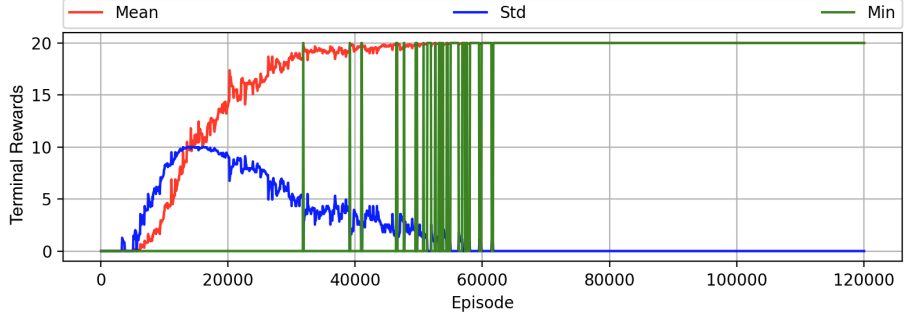

**Fig. 5   Optimization Terminal Reward History**

## IV. Experiments

Once the Meta-RL policy was optimized, we conducted multiple experiments to determine the performance of the Meta-RL guidance system and analyze its generalization capability. For the reader's convenience the labels and descriptions of all of the test scenarios we carried out are given in Table 6. We measure performance using three metrics. The first metric is the success rate, which we define as a miss distance less than 5 m and a terminal speed of at least 1700 m/s. The second metric is whether path constraints were satisfied. In order to independently assess both metrics, constraints are inactivated during testing in that they do not cause the episode to terminate. However, constraint violations are monitored.

**Table 6   Experiment Descriptions**

| Label | Description |
| --- | --- |
| Optim | Models parameter variation, actuator failure, actuator noise, and initial state/condition variation Tables 2 and 3, i.e., the same conditions used for optimization. |
| PV=$\Delta$ | Same as "Optim" experiment, but perturbs lift, drag, sideslip aerodynamic coefficients and atmospheric density by +/- $\Delta$% |
| MV/SV=$\Delta$ | Same as "Optim" experiment, but with the vehicle mass and reference area each (independently) randomly biased in the range +/-$\Delta$ at the start of each episode. |
| Divert=$\Delta$ | Same as "Optim" experiment, but divert distance set to $\Delta$% of range to target |
| Evasion=$\Delta$ | Same as "Optim" experiment, but multiple random diverts with divert distance set to $\Delta$% of range to target; last divert puts HSW on heading to actual Target |
| AF=0.5 | Branches off of the "Optim" experiment, but with $\epsilon_{\text{ctrl}} = (-0.5, 0.0)$. |

### A. Performance of Meta-RL Guidance System

To test the meta-RL guidance system, we ran 5000 episodes over the scenarios given in Table 6, with results tabulated in Table 7. The column labeled "Miss < 5m" and "Miss < 10m" indicate the percentage of episodes that resulted in a miss distance less than 5m and 10m, with a terminal speed of at least 1700 m/s. The column labeled "Violation" indicates the percentage of episodes that resulted in a constraint violation, with the "Type" column giving the violation type: load (Load), dynamic pressure (DP), or heating rate (HT). Performance under the "Optim" case is excellent, with all trajectories resulting in a miss distance less than 5m, and all path constraints satisfied. For the generalization cases, we see that performance degrades, but not catastrophically, and a few constraint violations begin to appear. The "Evasion" case induces rare load constraint violations; these are likely due to the required large course corrections. It is likely that performance over the generalization cases could be improved by considering these cases during optimization.

Table 8 gives statistics for the heating rate, load, and dynamic pressure constraints for the "Optim" case. Fig. 6 plots a histogram for time of flight and terminal speed, and Fig. 7 shows a miss distance scatter plot. The scatter plot was generated by re-running 5000 episodes, but terminating an episode when the altitude falls below zero rather than when the closing velocity turns negative. The scatter plot with the original data is squashed in the downrange direction, but has similar crossrange limits. We chose a random trajectory from the "Optim" case where a divert occurred at around 55 seconds into the trajectory. Fig. 8 illustrates the evolution of heating, dynamic pressure, and load for the sampled trajectory and Fig. 9 plots the sampled trajectory. Figure 10 plots the relative position between the HSW and target for 100 random trajectories in 3-D space. The discontinuities are due to the divert maneuvers, and all trajectories terminate at (0,0,0) because we are plotting relative position with respect to each target; in the actual engagement episode the terminal target positions were up to ten kilometers from the (0,0,0) position due to the impact of target motion and diverting to a new target. The combined effect of target motion and diverting to a new target is apparent in Fig. 11. Note that there appears to be a concentration of terminal target locations with a radius of around 2.2km. This is the extent of target motion at 30 m/s over the average flight time of 71 s.

Table 7    Performance

| Case | Miss (m) | | $\|V\|$ (m/s) | | Miss < 5m | Miss < 10m | Violation | Type |
|---|---|---|---|---|---|---|---|---|
| - | $\mu$ | $\sigma$ | $\mu$ | $\sigma$ | % | % | % | - |
| Optim | 1.4 | 0.8 | 2181 | 121 | 100.0 | 100.0 | 0.0 | - |
| PV=15 | 1.4 | 1.2 | 2179 | 152 | 99.4 | 99.6 | 0.0 | - |
| PV=20 | 1.7 | 3.3 | 2179 | 185 | 98.1 | 99.8 | 0.3 | DP |
| MV/SV=10% | 1.6 | 2.3 | 2178 | 144 | 98.9 | 99.4 | 0.1 | DP |
| AF=0.5 | 1.4 | 0.8 | 2181 | 121 | 99.9 | 100.0 | 0.0 | - |
| Divert=10% | 1.4 | 0.9 | 2181 | 124 | 99.9 | 100.0 | 0.3 | HT |
| Evasion=5% | 1.5 | 1.3 | 2183 | 124 | 99.9 | 99.5 | 0.2 | Load |

Table 8    Constraint Statistics ("Optim" Case)

| Constraint | $\mu$ | $\sigma$ | Max | Limit |
|---|---|---|---|---|
| Heating Rate (KW/s) | 4817 | 1554 | 8816 | 9000 |
| Load (m/s$^2$) | 21 | 11 | 104 | 147 |
| Dynamic Pressure (KPa) | 1070 | 889 | 3809 | 4000 |

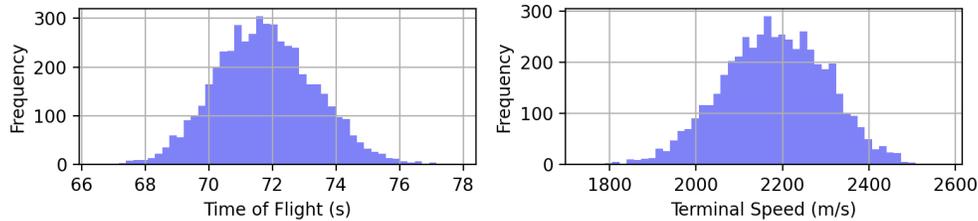

Fig. 6    Time of Flight and Terminal Speed ("Optim" Case)

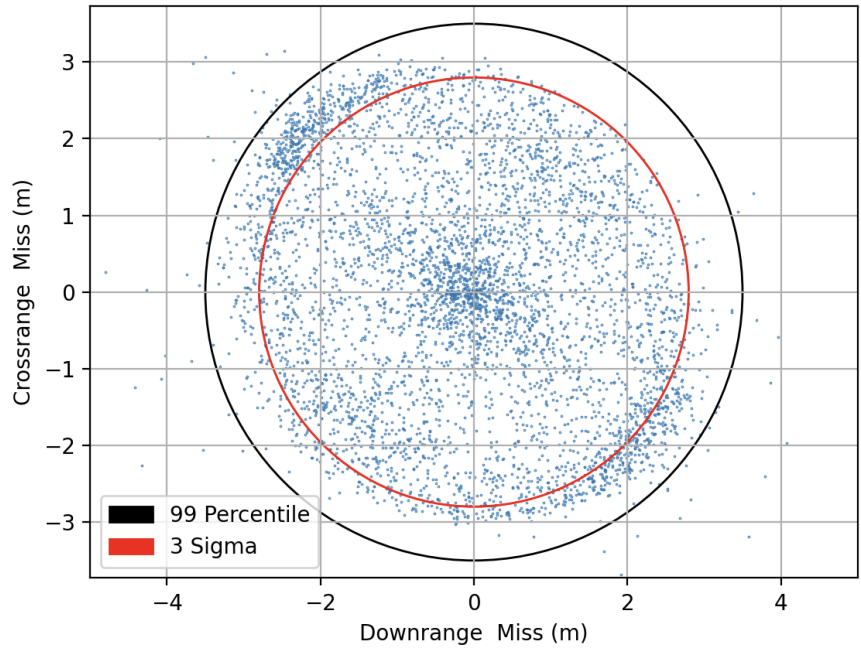

**Fig. 7    Accuracy ("Optim" Case)**

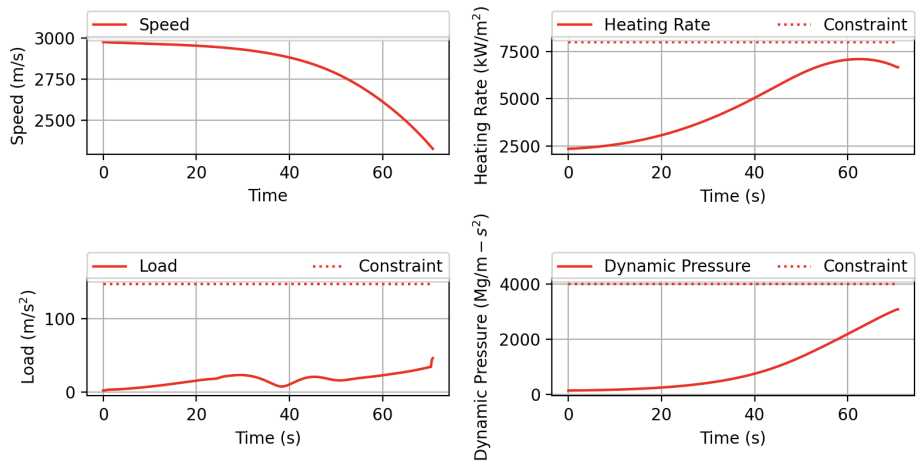

**Fig. 8    Sample Constraints ("Optim" Case)**

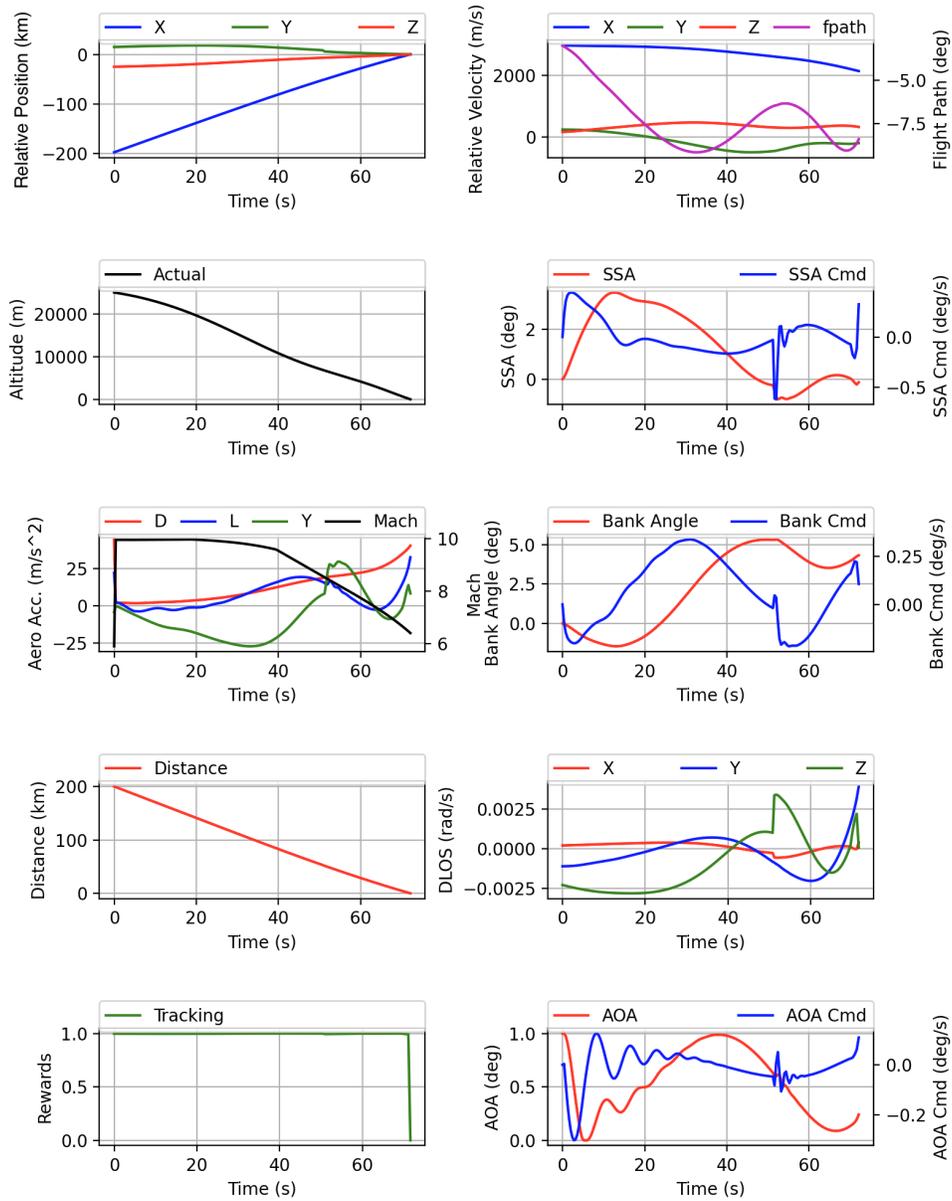

**Fig. 9  Sample Trajectory ("Optim" Case)**

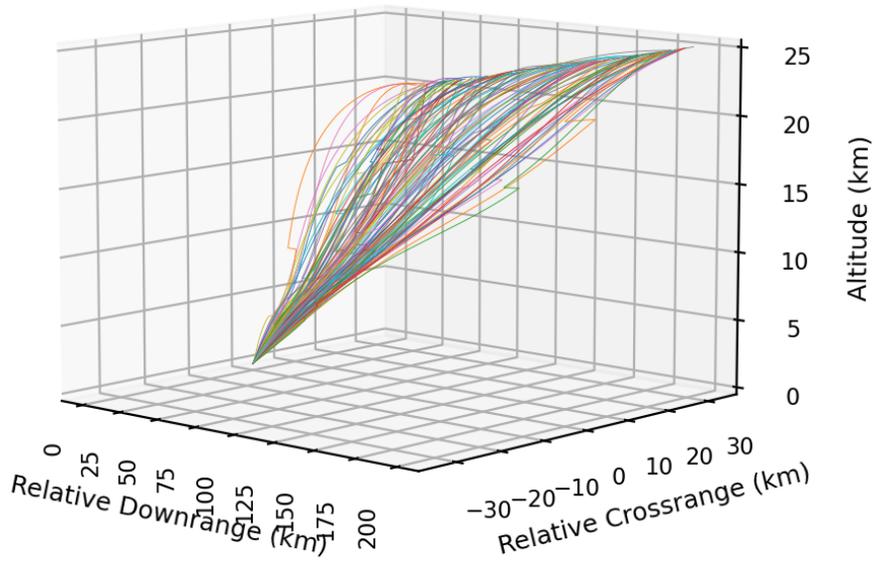

**Fig. 10    100 Trajectories ("Optim" Case)**

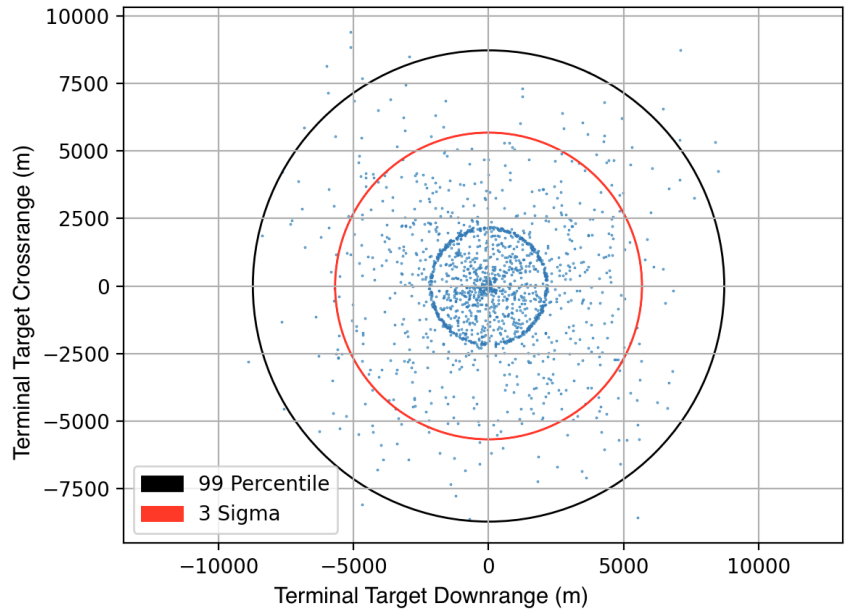

**Fig. 11    Target Dispersion ("Optim" Case)**

## B. Preliminary 6-DOF Results

The purpose of this experiment is to demonstrate that an integrated and adaptive guidance and flight control system can be optimized using meta-RL, and that the system does not need access to altitude, angle of attack, or sideslip angle measurements. Both [5] and [20] use the same vehicle geometry, but formulate the aerodynamic coefficients differently. We were originally going to use the aerodynamic model from [5], but found a large difference between the moment coefficients in [20] that were derived from computational fluid dynamics (CFD) simulation. For example, the total yawing moment with zero rudder and sideslip angle of zero versus angle of attack computed using the equations on page 13 of [20] was off by an order of magnitude when compared to the equation for $m_y$ in [5]. The equations in [20] were derived using the actual CFD data and have associated plots in [27] that we used to check our implementation, and have been used in multiple works by different authors. Thus, we used the aerodynamic coefficient models from [20], which we considered to be a more reliable source. It is worth noting that the aerodynamic force coefficients in [20] were in agreement with those in [5], except at higher Mach numbers than were used in this work.

We created a 6-DOF environment where the equations of motion are implemented as described in [17]. The randomized initial conditions are identical to that in our 3-DOF simulations, and we used the same reward function except that the control penalty was related to the commanded change in deflection angles. The observation space is shown below in Eq. (19a), where $\mathbf{q}$, $\omega$, $\delta_{LE}$, $\delta_{RE}$, and $\delta_{R}$, are the vehicle attitude, rotational velocity, left elevon angle, right elevon angle, and rudder angle, respectively. As with our 3-DOF implementation, all observations are readily available from the seeker and rate gyros. The action space is shown in Eqs. (20a) through Eqs. (20c), where $\Delta\delta_{LE}$, $\Delta\delta_{RE}$, $\Delta\delta_{RE}$ are the commanded change to the left elevon, right elevon, and rudder, respectively, and $\Delta\delta_{LE}^{max} = \Delta\delta_{RE}^{max} = \Delta\delta_{R}^{max} = 20°$. The policy output is $\mathbf{u} = \pi(\mathbf{o})$. The vehicle, which is the same used for our 3-DOF work, is shown in Fig. 12, which is reproduced from [20].

These results are preliminary, in that this is still work in progress, and performance is inferior to that of our 3-DOF results. We believe this might be due to the limited flight conditions used to build the aerodynamic model, particularly the angle of attack range. Future 6-DOF work for the HSW application will use a different vehicle geometry (closer to a lifting body such as HTV-2), with aerodynamics characterized at negative angles of attack. Clearly, the results shown in Table 9 do not match those in our 3-DOF results, although without divert maneuvers the performance improves. In future work we will close this gap. Interestingly, in 6-DOF the terminal speed has a tighter distribution, and heating rates are lower. Load and dynamic pressure are similar to the 3-DOF results.

$$\mathbf{o} = \begin{bmatrix} \lambda & \Omega & v_c & r & \mathbf{q} & \omega & \delta_{LE} & \delta_{RE} & \delta_R \end{bmatrix} \tag{19a}$$

$$\Delta\delta_{LE} = \text{clip}\left(\mathbf{u}_0 \Delta\delta_{LE}^{max}, -\Delta\delta_{LE}^{max}, \Delta\delta_{LE}^{max}\right) \tag{20a}$$
$$\Delta\delta_{RE} = \text{clip}\left(\mathbf{u}_1 \Delta\delta_{RE}^{max}, -\Delta\delta_{RE}^{max}, \Delta\delta_{RE}^{max}\right) \tag{20b}$$
$$\Delta\delta_{R} = \text{clip}\left(\mathbf{u}_2 \Delta\delta_{R}^{max}, -\Delta\delta_{R}^{max}, \Delta\delta_{R}^{max}\right) \tag{20c}$$

**Table 9   6-DOF Performance**

| Case | Miss (m) | | $\|V\|$ (m/s) | | Miss < 5m | Miss < 10m | Violation | Type |
|---|---|---|---|---|---|---|---|---|
| - | $\mu$ | $\sigma$ | $\mu$ | $\sigma$ | % | % | % | - |
| No-Divert | 1.6 | 0.6 | 2151 | 29 | 99.7 | 100.0 | 0.0 | - |
| Optim | 5.4 | 6.2 | 2152 | 40 | 65.9 | 83.7 | 0.0 | - |

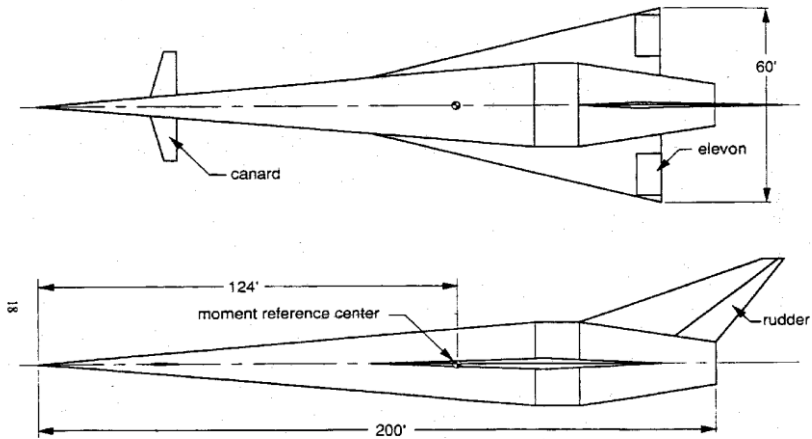

**Fig. 12   Generic Hypersonic Vehicle (GHV)**

## C. Discussion

There should not be any issues implementing the guidance policy on a flight computer, as although it can take several days to optimize a policy, the deployed policy can be run forward in a few milliseconds, as the forward pass consists of a few multiplications of small matrices. Although we did not include terminal flight path angle constraints, it would certainly be possible to shape the trajectory by including a terminal flight path limit in the terminal reward function. However, given the high speed and resultant kinetic energy, it is unlikely to make a difference in lethality except possibly for hardened land based targets, and we decided to focus on minimizing flight time. It is also possible that a shallow flight path (as demonstrated in this work) in contrast to a steeper trajectory may present difficulties for defensive systems designed to intercept ballistic missiles.

Note that the use of shaping rewards make the agent's behavior explainable, i.e., in this work the agent attempts to minimize the line of sight rotation rate while satisfying path constraints. Although Meta-RL provides a powerful framework for optimizing aerospace guidance, navigation, and control systems, there may be concern as to the optimized policy's deployment behavior when the agent is confronted with a scenario that differs substantially from the scenarios experienced during optimization. Our inclusion of generalization cases addresses this issue, but we also ran another experiment where the policy was optimized without divert maneuvers and then tested with divert maneuvers under the same conditions given in Table 6. From the perspective of the agent, a divert maneuver looks like the target instantly teleports to a new position and velocity. This certainly qualifies as a novel scenario far from anything experienced by the agent during optimization. However, performance was not impacted as compared to the results given in Section IV. Despite the novel target behavior, the policy generated actions that were consistent with the reward shaping function and path constraints.

# V. Conclusion

We optimized a guidance system for the terminal phase of a hypersonic strike weapon using reinforcement meta learning. The guidance system maps observations that include seeker angles and line of sight rotation rate to actions consisting of the commanded rates of change for bank, angle of attack, and sideslip angles. The optimization shaping reward function attempts to minimize the line of sight rotation rate, and the guidance system can be considered a form of proportional navigation that satisfies path constraints on heating rate, dynamic pressure, and load. Importantly, the observations used by the guidance law would be available from a radar seeker with minimal processing. The optimized system was tested over a range of scenarios inducing off-nominal flight conditions, including perturbation of aerodynamic coefficients, sensor noise, and actuator failure scenarios. We demonstrated that the guidance system is capable of both divert and evasive maneuvers, which are important capabilities for an autonomous hypersonic strike weapon. To our knowledge this is the first published work applying reinforcement meta learning to the optimization of a guidance system suitable for the terminal phase of a hypersonic strike weapon, and the first to consider divert and evasive maneuvers in this application. Future work will extend this work to six degrees-of-freedom with consideration of flexible modes, and explore multi-agent optimization for a group of hypersonic strike weapons.